# Can AB Dor C be a Tight Binary Brown Dwarf?

Christian Marois[1], Bruce Macintosh[1], Inseok Song[2] & Travis Barman[3]

[1] Institute of Geophysics and Planetary Physics L-413,
Lawrence Livermore National Laboratory, 7000 East Ave, Livermore, CA 94550

[2] Gemini Observatory, Northern Operations Center, Hilo, HI 96720

[3] Department of Physics and Astronomy, University of California at Los Angeles, Box 951547, Knudsen Hall, Los Angeles, CA 90095;

cmarois@igpp.ucllnl.org, bmac@igpp.ucllnl.org, song@gemini.edu & barman@astro.ucla.edu

AB Dor C, a low-mass companion to the young star AB Dor A has recently been imaged[1]. Direct detection and astrometric[2] observations have been used to solve the AB Dor C orbit and obtain a dynamical mass estimate. Near-infrared magnitudes and effective temperatures from models have then been compared with observed values. Models and observations disagree at more than the 2σ level. This result prompts Close et al. 2005 to claim that models are overestimating young low-mass object luminosities by roughly a factor of two. This claim is based on the hypothesis that the detected source, AB Dor C, is a single object. Another possible interpretation of the data that was not considered is that AB Dor C is an unresolved binary brown dwarf. Considering that 21 brown dwarfs companions have been found to date around stars and that three of them are confirmed or potential brown dwarf binaries (Gl569Ba & Bb, GL564B & C and Indi Ba & Bb)[3-5], this suggests that brown dwarf binaries may constitute a significant fraction (14% ± 8%) of the brown dwarf population around stars.

The binary brown dwarf hypothesis model is constructed assuming two brown dwarfs with total mass equal to the dynamical estimated mass 90 ± 10 $M_{Jup}$[1]. The derived spectral type (M8 ± 1)[1] of AB Dor C and the latest spectral type to effective temperature scale[6] imply an effective temperature of $2710^{+170}_{-310}$ K. The DUSTY model[7] for 2MASS photometric system (Baraffe & Allard, private communication) and estimated AB Dor system age (30-100 Myr) are used to calculate the predicted near-infrared magnitudes (*J*, *H* and *Ks*) and effective temperature profiles for different mass ratios of the two brown dwarfs (see Fig. 1). A mass ratio greater than ~0.3 can fit all observations within the 2σ observational error bars and the accepted age interval.

To be stable, the two brown dwarfs need to be gravitationally bound. From the derived orbit and estimated mass ratios, the periaston of AB Dor C is approximately 2 AU, so the separation between the two brown dwarfs must be less than ~0.5 AU and the orbital period is less than 500 days. At 14.9 pc and if viewed pole-on, the two brown dwarfs would be less than 34 mas apart at apoastron, < λ/D in *H* band on a diffracted limited 8-m telescope. Additionally, the mass ratio derived from observations and models show that one brown dwarf could be 30% of the mass of the other. Assuming a mass of

20 and 70 $M_{Jup}$ for the two brown dwarfs and an age between 30 and 100 Myr, models predict that the less massive brown dwarf could be 2 to 4 magnitudes fainter in *H* band and undetectable in Close et al.'s observations.

Radial velocity analysis or higher resolution images of AB Dor C could confirm its binarity. In order for AB Dor C to be yet another example of an object that disagrees with model predictions[8-10], the binary brown dwarf hypothesis must be ruled out.

**References**

1 Close L. M., Lenzen, R., Guirado, J. C., Nielsen, E. L., Mamajek, E. E., Brandner, W., Hartung, M., Lidman, C., Biller, B. *Nature* **433**, 287-289 (2005).
2 Guirado, J. C. et al. *Astrophys J.* **490**, 835-839 (1997).
3 Martìn, E. L., Koresko, C. D., Kulkarni, S. R., Lane, B. F., Wizinowich, P. L. *Astrophys. J.* **529**, L37-L40 (2000).
4 Goto, M., Kobayashi, N., Terada, H., Gaessler, W., Kanzawa, T., Takami, H., Takato, N., Hayano, Y., Kamata, Y., Iye, M., Saint-Jacques, D. J., Tokunaga, A. T., Potter, D., Cushing, M. *Astrophys. J.* **567**, L59-L62 (2002).
5 McCaughrean, M. J., Close, L. M., Scholz, R.-D., Lenzen, R., Biller, B., Brandner, W., Hartung, M., Lodieu, N. *Astron. Astrophys.* **413**, 1029-1036 (2004).
6 Luhman, K. L., Stauffer, J. R., Muench, A. A., Rieke, G. H., Lada, E. A., Bouvier, J., Lada, C. J., *Astrophys. J.* **593**, 1093-1115 (2003).
7 Chabrier, G., Baraffe, I., Allard, F., Hauschildt, P. *Astrophys. J.* **542**, 464-472 (2000).
8 Torres, G., Ribas, I. *Astrophys. J.* **567**, 1140-1165 (2002).
9 Mohanty, S., Jayawardhana, R., Basri, G. *Astrophys. J.* **609**, 885-905 (2004).
10 Zuckerman, B., Song, I. *ARA&A* **42**, 685-721 (2004).

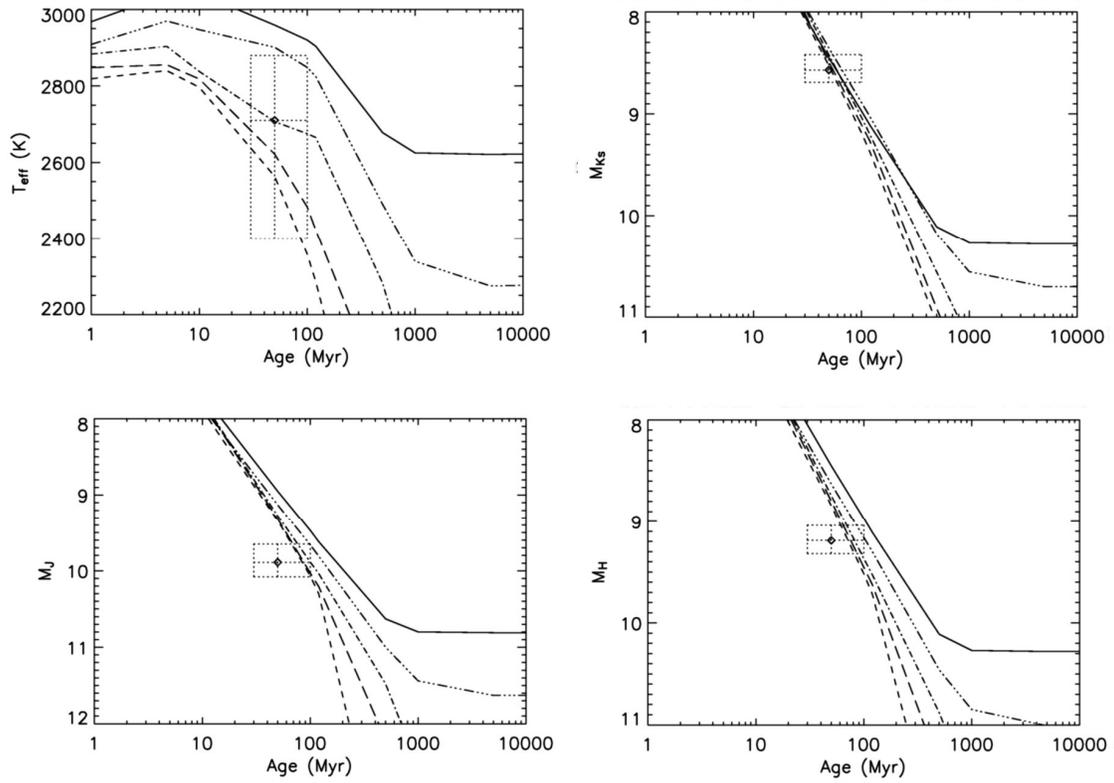

Fig. 1: Model predictions can explain the effective temperature and absolute *J, H* and *Ks* magnitudes for the accepted age interval if the binary brown dwarf hypothesis is assumed. The solid line shows the effective temperature/absolute magnitudes of a 90 $M_{Jup}$ object (Close et al. 2005 hypothesis). The dashed, long dashed, dot dashed and triple dots dashed lines show respectively the binary brown dwarf hypothesis with 90 $M_{Jup}$ total mass with mass ratios of 1, 0.66, 0.43 and 0.25. The diamond shows the AB Dor C observation data point with corresponding 2σ error bars[1].